\documentclass[11pt]{article}

\usepackage{latexsym}
\usepackage{amsthm}
\usepackage{amsmath,amssymb,amsfonts,bm,amsbsy,bbm,mathabx}
\usepackage{epsfig}
\usepackage{graphicx,rotating,lscape}
\usepackage{verbatim}
\usepackage{multirow,color}
\usepackage{geometry}
\usepackage{setspace}
\usepackage{float}
\usepackage{physics}
\usepackage{mathtools}
\usepackage{enumitem}
\usepackage{caption, subcaption}
\usepackage{authblk}

\usepackage{algorithm}
\usepackage{algorithmic}
\usepackage[algo2e]{algorithm2e}

\usepackage{natbib}

\DeclareMathOperator*{\argmin}{arg\,min}

\newtheorem{pro}{\underline{\bf Proposition}}
\newtheorem{Th}{\underline{\bf Theorem}}
\newtheorem{Rem}{\underline{\bf Remark}}

\newtheorem{Ass}{\underline{\bf Assumption}}

\def\bse{\begin{eqnarray*}}
\def\ese{\end{eqnarray*}}
\def\be{\begin{eqnarray}}
\def\ee{\end{eqnarray}}
\def\bsq{\begin{equation*}}
\def\esq{\end{equation*}}
\def\bq{\begin{equation}}
\def\eq{\end{equation}}
\def\bi{\begin{itemize}}
\def\ei{\end{itemize}}

\def\sumi{\sum_{i=1}^n}

\def\sumI{\sum_{i=1}^N}

\def\Theta{\theta}

\def\wh{\widehat}

\def\wt{\widetilde}

\def\pr{\hbox{pr}}

\def\calR{{\cal R}}

\def\trans{^{\rm T}}
\def\eff{_{\rm eff}}
\def\nv{_{\rm nv}}

\def\bb{{\boldsymbol\beta}}

\def\bg{{\boldsymbol\gamma}}

\def\bm{{\boldsymbol\mu}}

\def\bSigma{{\boldsymbol\Sigma}}

\def\ba{{\boldsymbol\alpha}}

\def\0{{\bf 0}}
\def\X{{\bf X}}
\def\x{{\bf x}}

\def\U{{\bf U}}

\def\b{{\bf b}}

\def\boxit#1{\vbox{\hrule\hbox{\vrule\kern6pt\vbox{\kern6pt#1\kern6pt}\kern6pt\vrule}\hrule}}

\def\expit{{\mbox{expit}}}

\def\calH{{\mathcal H}}
\def\calT{{\mathcal T}}

\def\fone{f_1}
\def\fzero{f_0}

\def\var{\hbox{Var}}

\makeatletter
\def\widebreve{\mathpalette\wide@breve}
\def\wide@breve#1#2{\sbox\z@{$#1#2$}%
     \mathop{\vbox{\m@th\ialign{##\crcr
\kern0.08em\brevefill#1{0.8\wd\z@}\crcr\noalign{\nointerlineskip}%
                    $\hss#1#2\hss$\crcr}}}\limits}
\def\brevefill#1#2{$\m@th\sbox\tw@{$#1($}%
  \hss\resizebox{#2}{\wd\tw@}{\rotatebox[origin=c]{90}{\upshape(}}\hss$}
\makeatletter

\begin{document}

\baselineskip 15 pt

\title{Incorporating External Controls for Estimating the Average Treatment Effect on the Treated with High-Dimensional Data: Retaining Double Robustness and Ensuring Double Safety
}

\author[a]{Chi-Shian Dai}
\author[b]{Chao Ying}
\author[c]{Yang Ning}
\author[b]{Jiwei Zhao \thanks{e-mail: \texttt{jiwei.zhao@wisc.edu}}}
\affil[a]{National Cheng Kung University}
\affil[b]{University of Wisconsin-Madison}
\affil[c]{Cornell University}

\maketitle

\begin{abstract}
Randomized controlled trials (RCTs) are widely regarded as the gold standard for causal inference in biomedical research.
For instance, when estimating the average treatment effect on the treated (ATT), a doubly robust estimation procedure can be applied, requiring either the propensity score model or the control outcome model to be correctly specified.
In this paper, we address scenarios where external control data, often with a much larger sample size, are available.
Such data are typically easier to obtain from historical records or third-party sources.
However, we find that incorporating external controls into the standard doubly robust estimator for ATT may paradoxically result in reduced efficiency compared to using the estimator without external controls.
This counterintuitive outcome suggests that the naive incorporation of external controls could be detrimental to estimation efficiency.
To resolve this issue, we propose a novel doubly robust estimator that guarantees higher efficiency than the standard approach without external controls, even under model misspecification.
When all models are correctly specified, this estimator aligns with the standard doubly robust estimator that incorporates external controls and achieves semiparametric efficiency.
The asymptotic theory developed in this work applies to high-dimensional confounder settings, which are increasingly common with the growing prevalence of electronic health record data.
We demonstrate the effectiveness of our methodology through extensive simulation studies and a real-world data application.
\end{abstract}


\newpage

\setcounter{equation}{0}

\section{Introduction}\label{sec:intro}
\noindent
Understanding the treatment effect is a fundamental and widely studied question in scientific research, whether in randomized controlled trials (RCTs) or observational studies.
Central to this inquiry are causal estimands \citep{Morgan_Winship_2014, imbens2015causal, hernan_robins_2022}, such as the Average Treatment Effect (ATE) and the Average Treatment Effect on the Treated (ATT), which provide insights into the overall and subgroup-specific impacts of interventions, respectively.
ATE quantifies the average treatment effect across the entire study population, whereas ATT focuses on the average effect among those who actually received the treatment.
As argued by \cite{heckmansmith1995} and \cite{heckman2001policy}, ATT is often the implicit target in many economics and evaluation studies, making it more relevant and interpretable for policymakers.
For example, when evaluating a government-implemented smoking cessation campaign, researchers and policymakers are likely more interested in the intervention's effect on those who actually participated (i.e., ATT) rather than on individuals who did not engage with the campaign \citep{wang2017g}.
A similar focus arises in medical research, where the primary interest often lies in assessing the treatment effect on actual participants.
Consider, for instance, a study investigating the causality between antidepressant use during pregnancy and the risk of autism spectrum disorder in children \citep{boukhris2016antidepressant}.
In such cases, the pertinent effect of interest is the impact of antidepressant use during pregnancy among women who have actually taken these medications. This subgroup likely includes individuals with symptoms of depression, anxiety, or other psychiatric disorders for which antidepressants are commonly prescribed.
The goal is to understand the average effect of administering antidepressants versus withholding them among women prescribed these medications, rather than the average effect across all women.

With advancements in modern technology and innovative data collection methods, researchers often have access to additional data sources, beyond the primary study which could be either a RCT or an observational study.
These external data, however, may not share the same structure as the primary study.
For instance, they might include only a control arm, as is often the case with data derived from historical records or third-party sources.
The integration of external control data raises intriguing statistical challenges regarding how to use these data safely and effectively.
In this paper, we adopt a potential outcomes framework within the primary study, focusing on estimating ATT in the presence of external controls, with potentially high-dimensional confounders.
We investigate when and how these external controls could improve ATT estimation, and propose methods for leveraging them effectively.

Our approach begins with the standard doubly robust estimator for ATT in the primary study, constructed using the efficient influence function.
This estimator is termed doubly robust because it requires only one of two models---either the propensity score model or the conditional mean model---to be correctly specified, though not necessarily both.
Building on this framework, a similar doubly robust estimator for ATT can be derived when incorporating external controls.
However, a key question arises: does the inclusion of external controls provide statistical benefits in terms of improved estimation efficiency?
For doubly robust estimators, this benefit is understood as follows: when only one of the two models (the propensity score model or the conditional mean model) is correctly specified, does the estimator that incorporates external controls remain more efficient than the one based solely on primary study data?
Existing research has explored the integration of external control data from both Bayesian \citep{yang2023sam, ohigashi2024nonparametric} and frequentist \citep{li2021improving, gao2023integrating, liao2023prognostic, schwartz2023harmonized, valancius2024causal, zhu2024enhancing} perspectives. However, ensuring the doubly safe use of external controls---guaranteeing efficiency gains while maintaining the doubly robust property---is a nontrivial challenge that has often been overlooked in the literature.

Another challenge addressed in this paper is the issue of high dimensionality, where the number of confounders can far exceed the sample size $n$ of the primary study.
Specifically, modeling high-dimensional propensity score or conditional mean models poses significant difficulties, whether with or without external controls.
The problem of managing high-dimensionality under model misspecification has garnered considerable attention in recent literature, including works by \cite{farrell2015robust}, \cite{belloni2018uniformly}, \cite{tan2020model}, and \cite{ghosh2022doubly}.
In this paper, we extend these insights to the framework of ATT estimation with external controls.
We show that high-dimensional nuisance models can be effectively handled to facilitate the development of a doubly robust inference procedure.

In summary, our methodology begins by proposing a doubly robust estimator $\wh\theta\eff$ based on the derived efficient influence function for estimating ATT in the presence of external controls.
This estimator is consistent when at least one of the nuisance working models is correctly specified and achieves the semiparametric efficiency bound when both models are correctly specified.
However, $\wh\theta\eff$ may exhibit lower efficiency than the naive doubly robust estimator $\wh\theta\nv$ which does not use external controls, in cases where only one of the nuisance working models is correctly specified.
This is problematic, as it implies that incorporating external controls could be detrimental to efficiency.
To address this issue, we propose an improved estimator, $\wh\theta_{\wh a}$, which guarantees an efficiency gain over the naive estimator $\wh\theta\nv$ even when only one nuisance working model is correctly specified.
We refer to this new estimator as doubly safe.
Additionally, both of our proposed estimator, $\wh\theta\eff$ and $\wh\theta_{\wh a}$, support a doubly robust inference procedure by effectively handling high-dimensional nuisance working models.

By doing so, our work makes two significant contributions.
Firstly, by appropriately addressing the high-dimensional nuisance working models and effectively applying the efficient influence function, our proposed estimators both $\wh\theta\eff$ and $\wh\theta_{\wh a}$ retain the double robustness property in both estimation and inference.
Secondly, the proposed estimator $\wh\theta_{\wh a}$ enjoys the doubly safe property, in that, even when only one nuisance working model is correctly specified, either the propensity score model or the conditional mean model, it still guarantees an efficiency gain over the naive estimator $\wh\theta\nv$ thus ensures the safe use of the external control data.

The rest of this paper is organized as follows.
In Section \ref{sec:notation}, we first introduce some preliminary knowledge of our framework that includes notation, assumptions, and parameter of interest.
Section~\ref{sec:methodology} is our main methodology, which includes the first proposed estimator $\wh\theta\eff$ in Section~\ref{sec:proposed1}, its comparison with the naive estimator $\wh\theta\nv$ in Section~\ref{sec:naive}, the second proposed estimator $\wh\theta_{\wh a}$ in Section~\ref{sec:method}, as well as the way of appropriately addressing the high-dimensional nuisance working models in Section~\ref{sec:nuisance}.
The corresponding theory is contained in Section~\ref{theory}.
Numerical experiments and a real data application are in Sections \ref{sec:simulation} and \ref{sec:realdata}, respectively.
The paper is concluded with some discussions in Section~\ref{sec:disc}.
All the technical proofs are deferred in the Supplement.

\section{Preliminaries}\label{sec:notation}


\subsection{Primary Study}

In applications, the primary study is usually a randomized controlled trial (RCT), but without loss of generality we consider a more general set of assumptions (details below) in the standard potential outcome framework \citep{neyman1923applications, rubin1974estimating}, where $Y_t$ is the outcome if the subject were assigned to treatment $T=t$, $t=0,1$.
Throughout the paper, we consider continuous outcomes, though our framework is able to be generalized to other outcome types.
We denote $\X$ the $(d+1)$-dimensional covariate with $1$ the first coordinate, and we also denote $n$ the sample size in the primary study.
We adopt the following standard assumptions:
\begin{itemize}
	\item[1] Consistency: The observed outcome $Y$ is equal to the potential outcome $Y_t$ when the treatment received is $t$, $t=0,1$. That is, the outcome we observe is $Y=Y_1 T + Y_0 (1-T)$.
	\item[2] Ignorability: The treatment assignment is conditionally independent of the potential outcomes given $\X$; i.e.,
	\be\label{eq:iginpri}
	T \perp \{Y_1, Y_0\} \mid \X.
	\ee
	In particular for RCT, we have $T \perp \{Y_1, Y_0, \X\}$.
	\item[3] Positivity: For each level of the covariate $\X$, the probability of receiving either treatment or control is greater than zero.

\end{itemize}

\subsection{External Controls, Notation, and Assumptions}

Besides the primary study, we assume that there exists $N-n$ external control subjects with $T=0$.
To facilitate the analysis with primary study and external controls combined, we introduce the binary variable $R$ in that $R=1$ represents the primary study and $R=0$ represents the external control.
For better illustration, the data structure of this paper is presented in Table~\ref{tb:data} below.
\begin{table}[!htbp]
	\caption{Data structure of this paper.}
	\centering
	\begin{tabular}{c|c|ccccc}
		\hline
		&   & $R$ & $T$ & $Y_1$ & $Y_0$ & $\X$\\
		\hline
		\multirow{4}{*}{Primary Study Subjects} & 1 & 1 & 1 & \checkmark & & \checkmark\\
		& 2 & 1 & 1 & \checkmark & & \checkmark\\
		& \vdots & 1 & 0 & & \checkmark & \checkmark\\
		& $n$ & 1 & 0 & & \checkmark & \checkmark\\
		\hline
		\multirow{3}{*}{External Control Subjects} & $n+1$ & 0 & 0 & & \checkmark & \checkmark \\
		& \vdots & 0 & 0 & & \checkmark & \checkmark \\
		& $N$ & 0 & 0 & & \checkmark & \checkmark \\
		\hline
	\end{tabular}\label{tb:data}
\end{table}

Throughout the paper, we use $\pr$ and $E$ to represent the probability and expectation of the population where all the $N$ subjects form a representative sample, and $\wt E$ the empirical average across the $N$ subjects.
For the relation between the primary study and the external controls, we assume
\be
&& \pr(T=1\mid \X,R=0)=0,\mbox{ and }\nonumber\\
&& \{R, T\} \perp \{Y_1, Y_0\} \mid \X.\label{eq:igincom}
\ee
Here, $\pr(T=1\mid \X,R=0)=0$ is self-explanatory, and $\{R, T\} \perp \{Y_1, Y_0\} \mid \X$ simply extends the ignorability from the primary study (\ref{eq:iginpri}) to the combined study (\ref{eq:igincom}).
From the ignorability in (\ref{eq:igincom}), one can write
\bse
&& \pr(T=1,R=1\mid y_1,y_0,\x)=\pr(T=1\mid \x)\\
&=& \underbrace{\pr(T=1,R=1\mid \x)}_{\delta(\x)}\\
&=& \underbrace{\pr(T=1\mid \x,R=1)}_{p(\x) } \underbrace{\pr(R=1\mid \x)}_{\pi(\x)},
\ese
where we define $\delta(\x)$, $p(\x)$ and $\pi(\x)$, respectively.
Clearly, $p(\x)$ encodes the propensity score in the primary study and equals the constant $p$ in a RCT, $\pi(\x)$ bridges the primary study and external controls formally, and $\delta(\x)=p(\x)\pi(\x)$.
We also denote $p=E\{p(\X)\mid R=1\}=\pr(T=1\mid R=1)$, $\pi=E\{\pi(\X)\}=\pr(R=1)\equiv \frac{n}{N}$, and $p\pi=\pr(R=1,T=1)=\pr(T=1)$.

We denote the marginal distribution of $\X$ in the primary study, $\X\mid R=1$, as $g(\x)$, and that in the external controls, $\X\mid R=0$, as $h(\x)$.
We allow $g(\x)\neq h(\x)$.
Clearly the marginal distribution of $\X$ in the combined study is $\pi g(\x)+ (1-\pi)h(\x)$, and we have $\pi(\x) = \frac{\pi g(\x)}{\pi g(\x)+ (1-\pi)h(\x)}$.
Finally, the ignorability (\ref{eq:igincom}) also indicates
\bse
R \perp Y_t \mid \X, t=0,1,
\ese
which means, the conditional distribution of $Y_t$ given $\X$ remains the same either $R=1$ or $R=0$.
For simplicity, we denote $f_t(y\mid \x)$ the conditional density/mass function of $Y_t$ given $\X$, and $\mu(\x)\equiv E(Y_0\mid\X=\x)$.

\subsection{Parameter of Interest}

In this paper, our parameter of interest is the ATT in the primary study. That is,
\be
\theta &=& E (Y_1-Y_0\mid T=1, R=1) \nonumber\\
&=& E (Y_1-Y_0\mid T=1) \label{eq:thetasame}\\
&=& \frac{\int\int y p(\x)\fone(y\mid\x)g(\x)\dd y \dd \x - \int\int y p(\x)\fzero(y\mid\x)g(\x)\dd y \dd \x}{\int p(\x)g(\x)\dd \x}.\label{eq:theta}
\ee
It is interesting to note that, from (\ref{eq:thetasame}), the ATT in the primary study is actually the same as the ATT in the combined study, because we do not have $T=1$ subjects in external controls.
Also, from (\ref{eq:theta}), the only term that external controls might be beneficial for estimation is $f_0(y\mid\x)$.
Indeed, the interesting phenomenon we figure out in this paper is, even though the combined study has a much larger sample size than the primary study for estimating $f_0(y\mid \x)$, it does not automatically indicate a guaranteed statistical benefit.
It heavily depends on how the statistical procedure, including both estimation and inference, is devised, especially under high-dimensionality.

\section{Methodology}\label{sec:methodology}

\subsection{Proposed Estimator with External Controls}\label{sec:proposed1}

Putting primary study data and external controls together, we consider the Hilbert space $\calH$ of all one-dimensional zero-mean measurable functions of the data following the likelihood function proportional to
\bse
p(\x)^{rt}\{1-p(\x)\}^{r(1-t)}\fone(y\mid\x)^{rt} \fzero(y\mid\x)^{1-rt}g(\x)^r h(\x)^{1-r},
\ese
with finite variance, equipped with the inner product $\langle h_1, h_2\rangle = E\{h_1(\cdot)h_2(\cdot)\}$ where $h_1(\cdot), h_2(\cdot)\in\calH$.
Our first result below gives an orthogonal decomposition of the semiparametric tangent space $\calT$ \citep{bickel1993efficient, tsiatis2006semiparametric} that is defined as the mean squared closure of the tangent spaces of parametric submodels spanned by the score vectors, and provides the efficient influence function for estimating $\theta$.
\begin{pro}\label{prop:eif}
	The semiparametric tangent space $\calT$ can be decomposed as
	\bse
	\calT = \Lambda_p \oplus \Lambda_1 \oplus \Lambda_0 \oplus \Lambda_{g} \oplus \Lambda_{h},
	\ese
	where
	\bse
	\Lambda_p &=& \left[r \ a(\x)\{t-p(\x)\}: \forall a(\x)\right],\\
	\Lambda_1 &=& \left\{ rt \ s_1(y,\x): \int s_1(y,\x)\fone(y\mid\x)\dd y=0 \right\},\\
	\Lambda_0 &=& \left\{ (1-rt) \ s_0(y,\x): \int s_0(y,\x)\fzero(y\mid\x)\dd y=0 \right\},\\
	\Lambda_{g} &=& \left[ r \ u(\x): \int u(\x)g(\x) \dd \x=E\{u(\X)\mid R=1\}=0 \right],\\
	\Lambda_{h} &=& \left[ (1-r) \ v(\x): \int v(\x)h(\x) \dd \x=E\{v(\X)\mid R=0\}=0 \right],
	\ese
	are the tangent spaces with respect to $p(\x)$, $\fone(y\mid\x)$, $\fzero(y\mid\x)$, $g(\x)$ and $h(\x)$, respectively.
	The notation $\oplus$ represents the direct sum of two spaces that are orthogonal to each other.
	Further, the efficient influence function for estimating $\theta$ is $\phi\eff(\mu^*, \delta^*)$, where
	\be\label{CEIF}
	\phi\eff(\mu, \delta)\equiv \frac{rt\{y-\mu(\x)-\theta^*\} }{\pi p } -\frac{(1-rt)\delta(\x)\{y-\mu(\x)\}}{\pi p\{1-\delta(\x)\}},
	\ee
	where $\mu^*(\x)$ and $\delta^*(\x)$ represent the true models of $E(Y_0\mid \x)$ and $\pr(T=1\mid \x)$, $\theta^*$ represents the true value of the estimand of interest.
\end{pro}
The proof of Proposition~\ref{prop:eif} can be found in Supplement~\ref{sec:app.space}.
In general, one can construct the doubly robust and semiparametrically efficient estimator for $\theta$, according to the efficient influence function (EIF) in Proposition~\ref{prop:eif}.
To do so, one needs to build statistical models for $\mu^*(\x)$ and $\delta^*(\x)$, respectively.
Under high dimensionality, bearing in mind the tradeoff between the model complexity and model interpretability, we fit working models $\mu\eff(\x)\equiv \x\trans\ba\eff$ and $\delta(\x)\equiv \expit(\x\trans\bg)$.
Using the combined data from both primary study and external controls, we denote the estimate of $\ba$ as $\wh\ba$, its asymptotic limit as $\overline{\ba}$, and similarly, $\wh\bg$ and $\overline\bg$, respectively.
If $\overline\mu\eff(\x)\equiv \x\trans\overline{\ba}$ equals $\mu^*(\x)$, we call the working model $\mu\eff(\x)=\x\trans\ba\eff$ correctly specified and otherwise misspecified.
Similarly, if $\overline\delta(\x)\equiv \expit(\x\trans\overline\bg)$ equals $\delta^*(\x)$, we call the working model $\delta(\x)=\expit(\x\trans\bg)$ correctly specified and otherwise misspecified.
Thus, our proposed estimator is
$\wh\theta\eff(\wh\ba\eff, \wh \bg)$, where
\be\label{eq:effDR}
\wh\theta\eff(\ba\eff, \bg) = \frac{\frac1N\sumI \left\{ r_i t_i (y_{i}-\x_i\trans\ba\eff)
	- (1-r_i t_i)\exp(\x_i\trans\bg) (y_{i}-\x_i\trans\ba\eff) \right\}
}{\frac1N\sumI r_i t_i
}.
\ee
For convenience, we also denote
\be\label{EFF}
\phi\eff(\ba\eff, \bg)
\equiv \frac{rt(y-\x\trans\ba\eff-\theta^*) }{\pi p } -\frac{(1-rt)\exp(\x\trans\bg)(y-\x\trans\ba\eff)}{\pi p}.
\ee

By appropriately devising nuisance parameter estimates $\wh\ba\eff$ and $\wh\bg$ to be introduced in Section~\ref{sec:nuisance}, Proposition \ref{prop:coef} in Section \ref{th:1} rigorously shows the desired theoretical properties of the proposed estimator $\wh\theta\eff(\wh\ba\eff, \wh \bg)$; i.e., double robustness for estimation as well as double robustness for inference \citep{zhang2021dynamic, athey2018approximate, sun2022high, ning2020robust, baybutt2023doubly, smucler2019unifying, wu2021model}.
Roughly, we have, for estimator $\wh\theta\eff(\wh\ba\eff, \wh \bg)$,
\bi
\item[DR1eff.] If $\overline\mu\eff=\mu^*$ but $\overline\delta\neq \delta^*$, it is consistent, and its variance is $E\{\phi^2\eff(\mu^*, \overline \delta)\}/N$;
\item[DR2eff.] If $\overline\delta=\delta^*$ but $\overline\mu\eff\neq\mu^*$, it is consistent, and its variance is $E\{\phi^2\eff(\overline\mu\eff, \delta^*)\}/N$;
\item[DR3eff.] If both $\overline\mu\eff=\mu^*$  and $\overline\delta= \delta^*$, it is consistent and semiparametrically efficient, and its variance is $E\{\phi^2\eff(\mu^*, \delta^*)/N\}$.
\ei

\begin{Rem}
	It warrants to emphasize that this type of result is not trivial.
	In DR1eff and DR2eff when one of the two working models is misspecified, the variance of the estimator needs to take into account the uncertainty of the misspecified model $\overline\delta(\cdot)$ or $\overline\mu\eff(\cdot)$; however, the above result indicates that the corresponding variance still preserves the same format as the situation when both working models are correctly specified.
\end{Rem}

\subsection{Naive Estimator without External Controls and the Comparison}\label{sec:naive}

In the absence of external controls, the naive estimator corresponds to the one that relies on the efficient influence function in the primary study.
This is a standard result in the literature.
Specifically, the EIF in \eqref{CEIF} degenerates into $\phi\nv(\mu^*,p^*)\equiv \frac{r}{\pi}\varphi\nv(\mu^*,p^*)$, where
\bse
\varphi\nv(\mu,p)&=& \frac{t\{y-\mu(\x)-\theta^*\} }{p } -\frac{(1-t)p(\x)\{y- \mu(\x)\}}{p\{1-p(\x)\}}.
\ese
Similar to Section~\ref{sec:proposed1}, we also fit working models $\mu\nv(\x)\equiv \x\trans\ba\nv$, $p(\x)\equiv \expit(\x\trans\bb)$, and denote
\be\label{eq:nvIF2}
\varphi\nv(\ba\nv,\bb)
= \frac{t(y-\x\trans\ba\nv-\theta^*) }{p } -\frac{(1-t)\exp(\x\trans\bb)(y- \x\trans\ba\nv)}{p}.
\ee
Thus, the naive estimator is $\wh\theta\nv(\wh\ba\nv, \wh\bb)$, where
\be\label{eq:nvDR}
\wh\theta\nv(\ba\nv,\bb) = \frac{\frac1n\sumi \left\{t_i (y_{i}- \x_i\trans\ba\nv) - (1-t_i) \exp(\x_i\trans\bb)(y_{i}- \x_i\trans\ba\nv)\right\}
}{\frac1n\sumi t_i
}.
\ee

By appropriately devising nuisance parameter estimates $\wh\ba\nv$ and $\wh\bb$ to be introduced in Section~\ref{sec:nuisance}, Proposition \ref{prop:coef2} in Supplement~\ref{th:2} rigorously shows the desired theoretical properties of the naive estimator $\wh\theta\nv(\wh\ba\nv, \wh \bb)$; i.e., double robustness for estimation as well as double robustness for inference.
Roughly, we have, for estimator $\wh\theta\nv(\wh\ba\nv, \wh \bb)$,
\bi
\item[DR1nv.] If $\overline\mu\nv=\mu^*$ but $\overline p\neq p^*$, it is consistent, and its variance is $E\{\varphi^2\nv(\mu^*, \overline p)\mid R=1\}/n$;
\item[DR2nv.] If $\overline p=p^*$ but $\overline\mu\nv\neq\mu^*$, it is consistent, and its variance is $E\{\varphi^2\nv(\overline\mu\nv, p^*)\mid R=1\}/n$;
\item[DR3nv.] If both $\overline\mu\nv=\mu^*$ and $\overline p= p^*$, it is consistent and semiparametrically efficient when only using primary study data, and its variance is $E\{\varphi^2\nv(\mu^*, p^*)\mid R=1\}/n$.
\ei

\begin{Rem}
	As a special case, when the primary study is a RCT, there is no misspecification issue of the model $p(\x)$. Then the naive estimator becomes $\wh\theta\nv(\wh\ba\nv)$ which is robust to the misspecification of the model $\mu\nv(\x)$, where
	\bse
	\wh\theta\nv(\ba\nv) = \frac{\frac1n\sumi \frac{t_i-p}{1-p} (y_{i}- \x_i\trans\ba\nv)
	}{\frac1n\sumi t_i
	}.
	\ese
\end{Rem}

Now we can evaluate the advantages of incorporating external control data by comparing the three properties of $\wh\theta\eff(\wh\ba\eff,\wh\bg)$ with $\wh\theta\nv(\wh\ba\nv,\wh\bb)$.
The comparison of DR3eff versus DR3nv is straightforward: when all of the four working models are correctly specified in that $\overline\mu\nv=\overline\mu\eff=\mu^*$, $\overline p= p^*$  and $\overline\delta= \delta^*$, the proposed estimator $\wh\theta\eff(\wh\ba\eff,\wh\bg)$ is certainly no less efficient than the naive estimator $\wh\theta\nv(\wh\ba\nv,\wh\bb)$.
However, such a guarantee may not be ensured under propensity score misspecification (DR1eff versus DR1nv) or outcome model misspecification (DR2eff versus DR2nv).
Specifically, we have
\begin{itemize}\label{lemma:twonaive}
	\item[(comp1)] DR1eff versus DR1nv: if $\overline\mu\nv=\overline\mu\eff=\mu^*$ but at least one of the two, $\overline p\neq p^*$ and $\overline\delta\neq \delta^*$, is noted, then $\wh\theta\eff(\wh\ba\eff,\wh\bg)$ may be less efficient than $\wh\theta\nv(\wh\ba\nv,\wh\bb)$; i.e., it is possible that $E\{\phi\eff^2(\mu^*,\overline \delta)\}>\pi^{-1} E\{\varphi\nv^2( \mu^*,\overline p)\mid R=1\}=E\{\phi\nv^2( \mu^*,\overline p)\}$. Note that this could happen even when the primary study is a RCT.
	\item[(comp2)] DR2eff versus DR2nv: if $\overline p=p^*$ and $\overline\delta=\delta^*$ but at least one of the two, $\overline\mu\nv\neq \mu^*$ and $\overline\mu\eff\neq \mu^*$, is noted, then, as long as $\overline\mu\nv \neq \overline\mu\eff$, the estimator $\wh\theta\eff(\wh\ba\eff,\wh\bg)$ may be less efficient than $\wh\theta\nv(\wh\ba\nv,\wh\bb)$; i.e., it is possible that $E\{\phi\eff(\overline\mu\eff ,\delta^*)^2\} > \pi^{-1} E\{\varphi\nv^2(\overline\mu\nv,p^*)\mid R=1\} = E\{\phi\nv^2(\overline\mu\nv,p^*)\}$.
	This could also happen when the primary study is a RCT.
\end{itemize}

Therefore, from either (comp1) or (comp2) above, directly employing the doubly robust estimator $\wh\theta\eff(\wh\ba\eff,\wh\bg)$ in practical applications may amplify the asymptotic variance thus is \underline{dangerous}.
In the following, we propose a novel estimator, that not only \underline{retains the double}\\
\underline{robustness property} in both estimation and inference as the estimator $\wh\theta\eff(\wh\ba\eff,\wh\bg)$, but also \underline{ensures the double safety property} in that it is guaranteed to be no less efficient than the naive estimator $\wh\theta\nv(\wh\ba\nv,\wh\bb)$ in scenarios (comp1) and (comp2) when some working models are misspecified.
Certainly when all the working models are correctly specified, this newly proposed estimator coincides the previously proposed estimator $\wh\theta\eff(\wh\ba\eff,\wh\bg)$, and both of them are semiparametrically efficient.

\subsection{Proposed Estimator that Ensures Double Safety}\label{sec:method}

The proposed estimator $\wh\theta_{\wh{a}}(\wh\ba\nv, \wh\ba\eff, \wh\bb, \wh\bg)$, shortly as $\wh\theta_{\wh a}$, satisfies the following three properties:
\bi
\item[DSDR1.] If $\overline\mu\nv=\overline\mu\eff=\mu^*$ but at least one of the two, $\overline p\neq p^*$ and $\overline\delta\neq \delta^*$, is noted, the estimator is consistent (doubly robust estimation), more efficient than $\wh\theta\nv(\wh\ba\nv,\wh\bb)$ (doubly safe estimation), and can produce asymptotically valid interval estimates (doubly robust inference);
\item[DSDR2.] If $\overline p= p^*$ and $\overline\delta=\delta^*$, but at least one of the two, $\overline\mu\nv\neq\mu^*$ and $\overline\mu\eff\neq\mu^*$, is noted, the estimator is consistent (doubly robust estimation), more efficient than $\wh\theta\nv(\wh\ba\nv,\wh\bb)$ (doubly safe estimation), and can produce asymptotically valid interval estimates (doubly robust inference);
\item[DSDR3.] If $\overline\mu\nv=\overline\mu\eff=\mu^*$, $\overline p= p^*$ and $\overline\delta=\delta^*$, the estimator is semiparametrically efficient, same as $\wh\theta\eff(\wh\ba\eff, \wh\bg)$.
\ei

The proposed estimator $\wh\theta_{\wh a}$ is defined as
\bse
\wh\theta_{\wh a}= \wh a \ \wh\theta\eff(\wh\ba\eff, \wh\bg) + (1-\wh a) \wh\theta\nv(\wh\ba\nv, \wh\bb),
\ese
where
\be\label{eq:estimatea}
\wh a = \frac{
	\frac1N \sumI \{\phi\nv(\wh\ba\nv,\wh\bb)-\phi\eff(\wh\ba\eff,\wh\bg)\}\phi\nv(\wh\ba\nv,\wh\bb)
}{
	\frac1N \sumI \{\phi\nv(\wh\ba\nv,\wh\bb)-\phi\eff(\wh\ba\eff,\wh\bg)\}^2
},
\ee
is an estimator of $a^*$, defined as
\be\label{eq:a}
a^*=\argmin_a \var\{a \ \wh\theta\eff(\overline\ba\eff, \overline\bg) + (1-a) \wh\theta\nv(\overline\ba\nv, \overline\bb)\}.
\ee
Theorem \ref{prop:hat-theta-a} in Section \ref{th:3} rigorously shows these desired theoretical properties of the proposed estimator $\wh\theta_{\wh a}$.

\subsection{Nuisance Parameter Estimation}\label{sec:nuisance}

In this section, we point out the basic ideas underlying the construction of the parameter estimators $\wh\ba\eff$, $\wh\ba\nv$, $\wh\bb$ and $\wh\bg$ such that the estimators $\wh\theta\eff(\wh\ba\eff, \wh\bg)$ and $\wh\theta\nv(\wh\ba\nv, \wh\bb)$ have the same large sample properties as $\wh\theta\eff(\overline\ba\eff, \overline\bg)$ and $\wh\theta\nv(\overline\ba\nv, \overline\bb)$, respectively, even with model misspecification.

Consider the Taylor expansion of $\wh\theta\eff(\wh\ba\eff, \wh\bg)$ and $\wh\theta\nv(\wh\ba\nv, \wh\bb)$:
\bse
\wh\theta\eff(\wh\ba\eff, \wh\bg)&=&\wh\theta\eff(\overline\ba\eff, \overline\bg)+\Delta_{11}+\Delta_{12}+o_p(N^{-1/2}),\\
\wh\theta\nv(\wh\ba\nv, \wh\bb)&=&\wh\theta\nv(\overline\ba\nv, \overline\bb)+\Delta_{21}+\Delta_{22}+o_p(N^{-1/2}),
\ese
with
\bse
\Delta_{11}&=&(\wh\ba\eff-\overline\ba\eff)\trans\times \frac{\partial}{\partial\ba}\wt{E}\{\phi\eff(\ba, \bg)\}\bigg|_{(\ba,\bg)=(\overline\ba\eff,\overline\bg)},\\
\Delta_{12}&=&(\wh\bg-\overline\bg)\trans\times \frac{\partial}{\partial\bg}\wt{E}\{\phi\eff(\ba, \bg)\}\bigg|_{(\ba,\bg)=(\overline\ba\eff,\overline\bg)},\\
\Delta_{21}&=&(\wh\ba\nv-\overline\ba\nv)\trans\times \frac{\partial}{\partial\ba}\wt{E}\{\phi\nv(\ba, \bb)\}\bigg|_{(\ba,\bg)=(\overline\ba\nv,\overline\bb)},\\
\Delta_{22}&=&(\wh\bb-\overline\bb)\trans\times \frac{\partial}{\partial\bb}\wt{E}\{\phi\nv(\ba, \bb)\}\bigg|_{(\ba,\bg)=(\overline\ba\nv,\overline\bb)},
\ese
where  the remainders are taken to be $o_p(N^{-1/2})$ under mild regularity conditions.

In the presence of model misspecification, note that $\Delta_{11}+\Delta_{12}+\Delta_{21}+\Delta_{22}$ is at least $O_p(N^{-1/2})$ because the second terms of $\Delta_{11}$, $\Delta_{12}$, $\Delta_{21}$ and $\Delta_{22}$ are typically of the order $O_p(1)$, e.g., using the regularized maximized likelihood estimator, and
the first terms are at least $O_p(N^{-1/2})$.
Therefore, in order that the estimators $\wh\theta\eff(\wh\ba\eff, \wh\bg)$ and $\wh\theta\nv(\wh\ba\nv, \wh\bb)$ possess desired theoretical properties,
we expect that all of the four terms $\Delta_{11}$, $\Delta_{12}$, $\Delta_{21}$ and $\Delta_{22}$ are of the order $o_p(N^{-1/2})$, which can be achieved if
the second terms in them are in the order of $o_p(1)$.
Thus, the population versions of the second terms in $\Delta_{11}$, $\Delta_{12}$, $\Delta_{21}$ and $\Delta_{22}$ are actually the estimating equations that the asymptotic limits $(\overline\ba\eff, \overline\bg, \overline\ba\nv, \overline\bb)$ are expected to satisfy:
\begin{equation}
	\label{eq:db}
	\begin{aligned}
		&& E\{(1-RT)(Y-\X\trans\overline\ba\eff)\exp(\X\trans\overline\bg)\X\} = \0,\\
		&& E\left[\{RT-(1-RT)\exp(\X\trans\overline\bg)\}\X\right] = \0,\\
		&& E\{R(1-T)(Y-\X\trans\overline\ba\nv)\exp(\X\trans\overline\bb)\X\} = \0,\\
		&& E\left[\{RT-R(1-T)\exp(\X\trans\overline\bb)\}\X\right] = \0.
	\end{aligned}
\end{equation}
This guides us the construction of the nuisance parameter estimates $\wh\ba\eff$, $\wh\bg$, $\wh\ba\nv$ and $\wh\bb$.
Specifically, to obtain $\wh\bg$ and $\wh\bb$, corresponding to the second and fourth equations in (\ref{eq:db}), we minimize the following Lasso penalized objective functions
\be
\ell_\lambda(\bg) &=& \frac{1}{N}\sumI \{(1-r_i t_i)\exp(\x_i\trans\bg)-r_i t_i \x_i\trans\bg\}+\lambda_{\bg} \|\bg_{-1}\|_1, \label{eq:estimatebg}\\
\ell_\lambda(\bb) &=& \frac{1}{N}\sumI \{r_i(1-t_i)\exp(\x_i\trans\bb)-r_i t_i \x_i\trans\bb\}+ \lambda_{\bb} \|\bb_{-1}\|_1, \label{eq:estimatebb}
\ee
where $\lambda_{\bg}$ and $\lambda_{\bb}$ are the tuning parameters, $\bg_{-1}=(\gamma_1,\cdots,\gamma_d)\trans$ excluding the intercept $\gamma_0$, $\bb_{-1}=(\beta_1,\cdots,\beta_d)\trans$ excluding the intercept $\beta_0$ and $\|\cdot\|_1$ denotes the $L_1$ norm.
Afterwards, the estimators $\wh\ba\eff$ and $\wh\ba\nv$ can be obtained by minimize the following Lasso penalized objective functions:
\be
\ell_{\lambda}(\ba\eff\mid \wh\bg) &=& \frac{1}{2N}\sumI
\{(1-r_i t_i)\exp(\x_i\trans\wh\bg)(y_i - \x_i\trans\ba\eff)^2\}+\lambda_{\ba\eff} \|\ba_{{\rm eff},-1}\|_1,\label{eq:estimatebaeff}\\
\ell_{\lambda}(\ba\nv\mid \wh\bb) &=& \frac{1}{2N}\sumI \{r_i(1-t_i)\exp(\x_i\trans\wh\bb)(y_i-\x_i\trans\ba\nv)^2\}+ \lambda_{\ba\nv} \|\ba_{{\rm nv}, -1}\|_1, \label{eq:estimatebanv}
\ee
where $\lambda_{\ba\eff}$ and $\lambda_{\ba\nv}$ are the tuning parameters, $\ba_{{\rm eff},-1}=(\alpha_{{\rm eff},1},\cdots,\alpha_{{\rm eff},d})\trans$ excluding the intercept $\alpha_{{\rm eff},0}$ and $\ba_{{\rm nv},-1}=(\alpha_{{\rm nv},1},\cdots,\alpha_{{\rm nv},d})\trans$ excluding the intercept $\alpha_{{\rm nv},0}$.

Now we briefly explain how to calculate the nuisance parameters in equations \eqref{eq:estimatebg}, \eqref{eq:estimatebb}, \eqref{eq:estimatebaeff} and \eqref{eq:estimatebanv}.
Firstly, in order to estimate $\bg$ and $\bb$, we adopt the Fisher scoring descent algorithm, which is studied in \cite{tan2020model} and can be obtained directly by using the \textbf{RCAL} package in \rm{R}.
After obtaining the estimators $\wh\bg$ and $\wh\bb$, insert them into the quadratic equations \eqref{eq:estimatebaeff} and \eqref{eq:estimatebanv}, which is a typical Lasso problem, so the \textbf{glmnet} package is directly used in \rm{R}.
Finally, we summarize the algorithm to compute the naive estimator $\wh\theta\nv(\wh\ba\nv,\wh\bb)$, the proposed estimator $\wh\theta\eff(\wh\ba\eff,\wh\bg)$, and the proposed estimator $\wh\theta_{\wh a}$ below in Algorithm~\ref{alg}.
\begin{algorithm}[tbp]
	\caption{Computing the estimators $\wh\theta\nv(\wh\ba\nv,\wh\bb)$, $\wh\theta\eff(\wh\ba\eff,\wh\bg)$ and $\wh\theta_{\wh a}$}
	\begin{algorithmic}
		\REQUIRE  Data $\{(R_i=1, T_i, Y_i, \X_i): i=1, \ldots, n\}$ from the primary study, and data $\{(R_i=0, T_i=0, Y_{0i}, \X_i): i=n+1, \ldots, N\}$ from external controls.
		\begin{itemize}
			\item[Step 1:] Using the primary study data only, compute $\wh\bb$ in (\ref{eq:estimatebb}) and $\wh\ba\nv$ in (\ref{eq:estimatebanv}), then compute $\wh\theta\nv(\wh\ba\nv, \wh\bb)$ where $\wh\theta\nv(\ba\nv,\bb)$ is defined in (\ref{eq:nvDR});
			\item[Step 2:] Using all the available data, compute $\wh\bg$ in (\ref{eq:estimatebg}) and $\wh\ba\eff$ in (\ref{eq:estimatebaeff}), then compute $\wh\theta\eff(\wh\ba\eff, \wh\bg)$ where $\wh\theta\eff(\ba\eff, \bg)$ is defined in (\ref{eq:effDR});
			\item[Step 3:] Compute $\wh a$ in (\ref{eq:estimatea}) where $\phi\nv$ is defined in (\ref{eq:nvIF2}) and $\phi\eff$ is defined in (\ref{EFF});
			\item[Step 4:] Compute $\wh\theta_{\wh a} = \wh a \ \wh\theta\eff(\wh\ba\eff, \wh\bg) + (1-\wh a) \wh\theta\nv(\wh\ba\nv, \wh\bb)$.
		\end{itemize}
	\end{algorithmic}\label{alg}
\end{algorithm}

\begin{Rem}\label{rem:comp2}
	The central idea of the method for estimating $\wh\ba\nv$, $\wh\ba\eff$, $\wh\bb$ and $\wh\bg$ proposed in Section~\ref{sec:nuisance} is to mitigate the impact of estimating these nuisance parameters on the estimation of the parameter of interest $\theta$, even in the presence of nuisance model misspecification.
	Specifically, the estimator $\wh\theta\eff(\wh\ba\eff, \wh\bg)$ (or, $\wh\theta\nv(\wh\ba\nv, \wh\bb)$) is designed to share the same asymptotic representation as $\wh\theta\eff(\overline\ba\eff, \overline\bg)$ (or, $\wh\theta\nv(\overline\ba\nv,\overline\bb)$).
\end{Rem}

\section{Theory}\label{theory}

\subsection{Theoretical Results for $\wh\theta\eff(\wh\ba\eff,\wh\bg)$ and $\wh\theta\nv(\wh\ba\nv,\wh\bb)$}\label{th:1}

The estimator $\wh\bg$ is the minimizer of $\ell_{\lambda}(\bg)$ in \eqref{eq:estimatebg}, which is a typical high-dimensional penalized lasso optimization problem, and the tuning parameter $\lambda_\bg$ in \eqref{eq:estimatebg} is usually specified as $\lambda_\bg=A_0\lambda_0$, with a constant $A_0>1$ and $\lambda_0 \gtrsim \sqrt{\log\ (d+1)/N}$. Further, the estimator $\wh\ba\eff$ is the minimizer of $\ell_{\lambda}(\ba\eff\mid \wh\bg)$, which is also a typical high-dimensional penalized lasso optimization problem, and the tuning parameter $\lambda_{\ba\eff}$ in \eqref{eq:estimatebaeff} is usually specified as $\lambda_{\ba\eff}=A_1\lambda_1$, with a constant $A_1>1$ and $\lambda_1 \gtrsim \sqrt{\log\ (d+1)/N}$.
For simplicity, we only present the most essential results for $\wh\bg$ and $\wh\ba\eff$ here, with more technical details deferred in Supplement~\ref{sec:app.prelim}.
To proceed, we need the following assumption:
\begin{Ass}\label{ass1}
	We assume
	\begin{itemize}
		\item[(i)] $\max_{j=1,\dots,d} |X_j|\leq C_0$, for a constant $C_0>0$.
		\item[(ii)]
		$\X\trans\overline\bg\leq B_0$ almost surely for $B_0\in\calR$; that is, $\overline\delta(\X)=\expit(\X\trans\overline\bg)$ is bounded from above by $(1+e^{-B_0})^{-1}.$
		\item[(iii)]
		The compatibility condition (defined below) holds for $\Sigma_\bg$ with subset $S_{\bg}=\{0\}\cup\{j:\gamma_j\ne 0, j=1,\dots, d\}$ and constants $\nu_0$ and $\xi_0>1$; it also holds for $\Sigma_\bg$ with subset $S_{\ba\eff}=\{0\}\cup\{j:\alpha_{{\rm eff},j}\ne 0, j=1,\dots, d\}$ and constants $\nu_1$ and $\xi_1>1$, where $\Sigma_\bg=E\{\exp(\X\trans\overline\bg)(1-RT) \X\X\trans\}$.
		\item[(iv)]
		$Y-\X\trans \overline\ba\eff$ is uniformly sub-Gaussian given $\X$: $D_0^2E[\{\exp(Y-\X\trans\overline\ba\eff)^2/D_0^2\}-1|\X  ]\leq D_1^2.$
		\item[(v)]
		$|S_\bg|\lambda_0=o(1)$ and $|S_{\ba\eff}|\lambda_1=o(1)$.
	\end{itemize}
\end{Ass}

Assumption~\ref{ass1}(i)-(ii), commonly used in the literature \citep{tan2020model, tan2020regularized}, is mainly used to bound the curvature of a symmetrized Bregman divergence associated with a non-quadratic loss function.

In Assumption~\ref{ass1}(iii), following the literature \citep{buhlmann2011statistics}, the compatibility condition holds for the $(d+1)\times(d+1)$ matrix $\Sigma$ with the subset $S\subset \{0,1,2,\dots,d\}$  and constants  $\nu_0$ and $\xi_0>1$,
if and only if
$$\nu^2_0\left(\sum_{j\in S}|b_j|\right)^2\leq |S|(\b\trans \Sigma\b)$$
for any vector $\b=(b_0,\dots,b_d)\trans\in\calR^{d+1}$ satisfying
$\sum_{j\notin S}|b_j|\leq \xi_0 \sum_{j\in S}|b_j|$, where $|S|$ denotes the cardinality of the set $S$.
By the Cauchy-Schwartz inequality, this condition is implied by the restricted eigenvalue (RE) condition that is popularly used in the high-dimensional statistics literature.
Assumption~\ref{ass1}(iii) can be generally verified if given the concrete form of the Gram matrix $E(\X\X\trans)$.
For instance, when $E(\X\X\trans)$ is an AR(1) matrix, $\Sigma_\bg$ satisfies the compatibility condition, in conjunction with some additional condition; e.g., for some constant $\tau_0>0$, $\b\trans E(\X\X\trans)\b\leq (\b\trans \bSigma_{\bg}\b)/\tau_0, \forall\ \b\in \calR^{d+1}$.
The matrix $\bSigma_{\bg}$ also equals $E\{\partial^2\ell_{\lambda}(\ba\eff\mid \wh\bg)/\partial\ba\eff\partial\ba\eff\trans\}$ evaluated at $\ba\eff=\overline{\ba}\eff$.

Assumption \ref{ass1}(iv) includes the case where the residual $Y-\X\trans\overline{\ba}\eff$ may not have mean 0 conditional on $\X$.
Assumption \ref{ass1}(v) is used to facilitate both the derivation of the empirical compatibility condition and the localized analysis with a non-quadratic loss function.

In what follows, we are able to present the convergence rates of $\wh\bg$ and $\wh\ba\eff$ in their $L_1$ norms, $\|\wh\bg-\overline\bg\|_1$ and $\|\wh\ba\eff-\overline\ba\eff \|_1$, respectively, and its proof can be found in Supplement~\ref{sec:app.prelim}.
\begin{pro}\label{prop:coef}
	Under Assumption \ref{ass1}, we have
	\bse\label{eq:gamma conv}
	\|\wh\bg-\overline\bg\|_1=O_p(|S_\bg|\lambda_0),  \mbox{ and }\|\wh\ba\eff-\overline\ba\eff \|_1=O_p(|S_\bg|\lambda^2_0\lambda_1^
	{-1}+|S_{\ba\eff}|\lambda_1).
	\ese
\end{pro}
\begin{Rem}[Obtaining sharp convergence rate]\label{sharprate}
	In the second result of Proposition~\ref{prop:coef}, the convergence rate $O_p(|S_\gamma|\lambda_0+|S_{\ba\eff}|\lambda_1)$ applies to any $\lambda_0 \gtrsim \sqrt{\log\ (d+1)/N}$ and $\lambda_1 \gtrsim \sqrt{\log\ (d+1)/N}$.
	From some simple derivations, when $\lambda_1 \asymp \lambda_0 \asymp \sqrt{\log\ (d+1)/N}$, it achieves the sharp convergence rate in that
	\bse
	\|\wh\ba\eff-\overline\ba\eff \|_1=O_p\{(|S_\bg|+|S_{\ba\eff}|)\lambda_0\}.
	\ese
\end{Rem}

The following result gives the convergence rate of $\wh\theta\eff(\wh\ba\eff,\wh\bg)$,
allowing both nuisance models to be misspecified, and its proof can be found in Supplement~\ref{wf}.
\begin{pro}\label{combine}
	Under Assumption \ref{ass1}, we have
	\bse\label{h1}
	|\wh\theta\eff(\wh\ba\eff,\wh\bg)-\wh\theta\eff(\overline\ba\eff,\overline\bg) |=O_p\left\{\pi^{-1} (|S_{\ba\eff}|+|S_\gamma|)\lambda_0\lambda_1 +\pi^{-1} |S_\gamma|\lambda_0^2\right\}.
	\ese
\end{pro}
\begin{Rem}\label{remark:eff}
	To achieve the sharp convergence rate of $|\wh\theta\eff(\wh\ba\eff,\wh\bg)-\wh\theta\eff(\overline\ba\eff,\overline\bg)|$, note that when $\lambda_1 \asymp \lambda_0 \asymp \sqrt{\log\ (d+1)/N}$, it obtains the rate $O_p\left\{\pi^{-1} (|S_{\ba\eff}|+|S_\bg|)\lambda_0^2\right\}$. Proposition
	\ref{combine} shows that $\wh\theta\eff(\wh\ba\eff,\wh\bg)$ is doubly robust for $\theta^*$ provided $\pi^{-1} (|S_{\ba\eff}|+|S_\bg|)\lambda_0^2=o(1)$. In addition, the following theorem gives the central limit theorem provided $(|S_{\ba\eff}|+|S_\bg|)\log\ (d+1)/n^{1/2}=o(1)$.
\end{Rem}
\begin{Th}\label{pro11}
	Under Assumption \ref{ass1}, if either of the two working models is correctly specified and $(|S_{\ba\eff}|+|S_\bg|)\log\ (d+1)/n^{1/2}=o(1)$, we have
	\bse
	\sqrt{N}\{\wh\theta\eff(\wh\ba\eff,\wh\bg)-\theta^*\}\xrightarrow{d} N(0,V\eff),
	\ese
	which can also be written as
	\bse
	\sqrt{n}\{\wh\theta\eff(\wh\ba\eff,\wh\bg)-\theta^*\}\xrightarrow{d} N(0,\pi V\eff),
	\ese
	where $V\eff =E\{\phi^2\eff(\overline\ba\eff,\overline\bg)\}$. Further, the $1-\alpha$ confidence interval of $\theta^*$ is
	\bse
	\left[\wh\theta\eff(\wh\ba\eff,\wh\bg)-z_{\alpha/2}\sqrt{\wh V\eff/N}, \wh\theta\eff(\wh\ba\eff,\wh\bg)+z_{\alpha/2}\sqrt{\wh V\eff/N}\right],
	\ese
	where $z_{\alpha/2}$ is the $1-\alpha/2$ quantile of $\mathcal{N}(0, 1)$ and $\wh V\eff=\wt{E}\{\phi\eff^2(\wh{\ba}\eff),\wh{\bg}\}$.
\end{Th}
Theorem~\ref{pro11} demonstrates that our confidence interval is doubly robust; that is, if either $\overline{\bg}=\bg^*$ or $\overline{\ba}\eff=\ba^*$, our confidence interval presented above is valid.
Theorem~\ref{pro11} also indicates that, when both working models are correctly specified; i.e., $\overline{\bg}=\bg^*$ and $\overline{\ba}\eff=\ba^*$,
the asymptotic variance of $\wh\theta\eff(\wh\ba\eff,\wh\bg)$ equals $E\{\phi\eff^2(\ba^*,\bg^*)\}/N$ which is the semiparametric efficiency bound for estimating $\theta^*$.

The theoretical results for $\wh\theta\nv(\wh\ba\nv,\wh\bb)$ can be similarly derived.
In the interest of space, we omit the presentation here and one can find the details in Supplement~\ref{th:2} and Supplement~\ref{pro:2}.

\subsection{Theoretical Results for $\wh\theta_{\wh{a}}$}\label{th:3}

Recall the definitions of $a^*$ and $\wh a$ in \eqref{eq:estimatea} and \eqref{eq:a}, respectively.
Below, we provide the convergence rate of the parameter estimator $\wh{a}$, and its proof can be found in Section \ref{sec:app.ahatandastar}.
\begin{pro}\label{lem:ahat}
	Under Assumptions \ref{ass1} and \ref{ass2} (presented in Supplement~\ref{th:2}), we have
	\bse\label{eq:ahat rate}
	|\wh{a}-a^*|= \pi^{1/2}O_p(|S_\bg|\lambda_0+|S_{\ba\eff}|\lambda_1+|S_\bb|\lambda_{01}+|S_{\ba\nv}|\lambda_{11}).
	\ese
\end{pro}

Proposition \ref{lem:ahat} shows that the convergence rate of our estimator $\wh{a}$ is dictated by the convergence rates of the four nuisance parameters. Furthermore, obtaining the convergence rate is beneficial for analyzing the estimator $\wh\theta_{\wh a}$, and its proof can be found in Section \ref{new}.

	\begin{pro}\label{prop:A new}
		Under Assumptions \ref{ass1} and \ref{ass2} (presented in Supplement~\ref{th:2}), we have
		\bse
		|\wh \theta_{\wh a}-\overline\theta_{a^*}|
		&=&			 O_p(1)			|a^*|\pi^{-1}\left\{(|S_{\ba\eff}|+|S_\bg|)\lambda_{0}\lambda_{1}+|S_\bg|\lambda_{0}^2\right\}\\
		&&+O_p(1)|1-a^*|
		\left\{(|S_{\alpha\nv}|+|S_\beta|)\lambda_{01}\lambda_{11} +|S_\beta|\lambda_{01}^2\right\}\\
		&&+\ O_p(1/\sqrt{N})(|S_\gamma|\lambda_0+|S_{\ba\eff}|\lambda_1+|S_\beta|\lambda_{01}+|S_{\ba\nv}|\lambda_{11}),
		\ese
		where $\overline\theta_{a^*}=a^* \wh\theta\eff(\overline\ba\eff, \overline\bg) + (1-a^*) \wh\theta\nv(\overline\ba\nv, \overline\bb)$.
	\end{pro}

	The results in Proposition~\ref{prop:A new} also indicate, when $a^*=1$, the convergence rate of $|\wh \theta_{\wh a}-\overline\theta_{a^*}|$ is $
	O_p\{\pi^{-1}(|S_{\ba\eff}|+|S_\bg|)\lambda_{0}\lambda_{1}\}+O_p(\pi^{-1}|S_\bg|\lambda_{0}^2)$, echoes that of the estimator $|\wh\theta\eff(\wh\ba\eff,\wh\bg)-\wh\theta\eff(\overline\ba\eff,\overline\bg) |$, while when $a^*=0$, the convergence rate of $|\wh \theta_{\wh a}-\overline\theta_{a^*}|$, \\ $O_p\left\{(|S_{\alpha\nv}|+|S_\beta|)\lambda_{01}\lambda_{11} +|S_\beta|\lambda_{01}^2\right\}$, reflects the convergence rate of $|\wh\theta\nv(\wh\ba\nv,\wh\bb)-\wh\theta\nv(\overline\ba\nv,\overline\bb) |$.
	
	Finally we present the asymptotic normality of the estimator $\wh\theta_{\wh a}$.
	
	\begin{Th}\label{prop:hat-theta-a}
		Assume Assumptions \ref{ass1} and \ref{ass2} (presented in Supplement~\ref{th:2}) hold, $(|S_{\ba\nv}|+|S_{\bb}|+|S_{\ba\eff}|+|S_{\bg}|)\log\ (d+1)/\sqrt{n}=o(1)$ and either of the two working models is correctly specified. Then, we have
		\bse
		\sqrt{N}(\wh\theta_{\wh a}-\theta^*)\xrightarrow{d} N(0,V_a),
		\ese
		which can also be written as
		\bse
		\sqrt{n}(\wh\theta_{\wh a}-\theta^*)\xrightarrow{d} N(0,\pi V_a),
		\ese
		where
		\bse
		V_a= E\{\phi\nv^2(\overline\ba\nv,\overline\bb)\} - \frac{
			\left(E\left[\{\phi\nv(\overline\ba\nv,\overline\bb)-\phi\eff(\overline\ba\eff,\overline\bg)\}\phi\eff(\overline\ba\eff,\overline\bg)\right]\right)^2
		}{E\left[\{\phi\nv(\overline\ba\nv,\overline\bb)-\phi\eff(\overline\ba\eff,\overline\bg)\}^2\right]}.
		\ese
		Further, the $1-\alpha$ confidence interval of $\theta^*$ is
		\bse
		\left[\wh\theta_{\wh a}-z_{\alpha/2}\sqrt{\wh V_a/N}, \wh\theta_{\wh a}+z_{\alpha/2}\sqrt{\wh V_a/N}\right],
		\ese
		where $z_{\alpha/2}$ is the $1-\alpha/2$ quantile of $\mathcal{N}(0, 1)$ and
		\bse
		\wh V_a=\wt{E}\{\phi\nv^2(\wh{\ba}\nv,\wh{\bb})\}-\frac{
			\left(\wt{E}\left[\{\phi\nv(\wh\ba\nv,\wh\bb)-\phi\eff(\wh\ba\eff,\wh\bg)\}\phi\eff(\wh\ba\eff,\wh\bg)\right]\right)^2
		}{\wt{E}\left[\{\phi\nv(\wh\ba\nv,\wh\bb)-\phi\eff(\wh\ba\eff,\wh\bg)\}^2\right]}.
		\ese 
	\end{Th}

	Theorem~\ref{prop:hat-theta-a} indicates that the estimator $\wh\theta_{\wh{a}}$ is no less efficient than the primary estimator $\wh\theta\nv$.
	By rewriting
	\bse
	V_a = E\{\phi\eff^2(\overline\ba\eff,\overline\bg)\} - \frac{
		E^2\left[\{\phi\nv(\overline\ba\nv,\overline\bb)-\phi\eff(\overline\ba\eff,\overline\bg)\}\phi\nv(\overline\ba\eff,\overline\bg)\right]
	}{
		E\left[\{\phi\nv(\overline\ba\nv,\overline\bb)-\phi\eff(\overline\ba\eff,\overline\bg)\}^2\right]
	},
	\ese
	it is also no less efficient than the estimator $\wh\theta\eff$.
	When both working models are correctly specified, we can obtain $a^*=1$, which means that the estimator $\wh\theta_{\wh{a}}=\wh\theta\eff$ is also an efficient estimator and achieves the semiparametric efficiency bound.
	
	\begin{Rem}[Discussion when $N\gg n$]
		Sometimes, the external controls might have a much larger sample size than the primary study, we thus consider the $N\gg n$ case.
		Interestingly, the much larger external controls' size does not help to reduce the convergence rate of the proposed estimators $\wh\theta\eff(\wh\ba\eff,\wh\bg)$ and $\wh\theta_{\wh a}$; i.e., their convergence rates remain to be $n^{-1/2}$, as indicated in Theorem~\ref{pro11} and Theorem~\ref{prop:hat-theta-a}.
		For the estimator $\wh\theta\eff(\wh\ba\eff, \wh\bg)$, one can easily check that $\pi V_1 =\pi E\{\phi^2\eff(\overline\ba\eff,\overline\bg)\} = O(1)$; thus, we write $\sqrt{n}\{\wh\theta\eff(\wh\ba\eff,\wh\bg)-\theta^*\}\xrightarrow{d} N(0,\pi V_1)$, as presented in Theorem~\ref{pro11}.
		This also applies to the estimator $\wh\theta_{\wh a}$.
	\end{Rem}

	\section{Simulations Studies}\label{sec:simulation}
	
	In this section, to comprehensively demonstrate the different performances of the three estimators $\wh\theta\nv$, $\wh\theta\eff$ and $\wh\theta_{\wh{a}}$,
	we consider the following two data generation models, the second of which considers three scenarios.
	
	For \underline{Model 1}, we first generate a $d$-dimensional multivariate normal random vector $\U \sim \mathcal{N}(\0,\Sigma)$ with $\Sigma_{jk}=2^{-|j-k|}$, $1\leq j,k\leq d$.
	Then we set the covariate $\X=(1,X_1,\dots,X_d)\trans$ to be $X_j=U_j ~\mathrm{for}~ 1\leq j\leq d$ and $\X^\dagger=(1,X^\dagger_1,\dots,X^\dagger_d)\trans$
	where $X^\dagger_j$ is the normalized version of $U_j+\{(U_j+1)_+\}^2$ with mean 0 and variance 1 for $j=1,\dots, 4$, and $X^\dagger_j=U_j ~\mathrm{for}~ 5\leq j\leq d$.
	Then, we generate the binary indicator $R$ by
	\bse
	\pi(\x)=\pr(R=1\mid \x) = \delta^*(\x)/p^*(\x),
	\ese
	where $p^*(\x)=\expit(0.125 x_1)$ and $\delta^*(\x)=\expit(-2+0.125 x_1)$.
	Further, for the subjects with $R=1$, we generate the treatment variable $T$ by
	\bse
	p(\x) = \pr(T=1\mid \x,R=1) = p^*(\x).
	\ese
	Finally, for the potential outcomes $Y_0$ and $Y_1$, we generate
	$Y_0=\mu^*_0(\X^\dagger)+\epsilon_0$ and
	$Y_1=\epsilon_1$, where $\epsilon_0\sim N(0,1)$, $\epsilon_1\sim N(0,1/2)$, and $\mu^*_0(\x)=x_1+0.5x_2+0.25x_3+0.125x_4$.
	Clearly, \underline{Model 1} is designed for the situation that, both models $p(\x)$ and $\delta(\x)$ will be correctly specified, while the outcome mean model $\mu_0(\x)$ will be misspecified.
	
	For \underline{Model 2}, we first generate a $d$-dimensional multivariate normal random vector $\U \sim \mathcal{N}(\bm,\Sigma)$ with  $\Sigma_{jk}=2^{-|j-k|}$, $1\leq j,k\leq d$, and set the covariate $\X=(1,X_1,\dots,X_d)\trans$ to be $X_j=U_j ~\mathrm{for}~ 1\leq j\leq d$.
	For subjects in the primary study with $R=1$, we set $\bm=\0$, while for external controls with $R=0$, we set $\mu_i=-3$ for $i=1\leq i\leq 4$ and $\mu_i=0$ for $i=5,\dots,d$.
	Equivalently, this indicates that
	\bse
	\pi(\x)=\pr(R=1\mid \x) = \expit\{\log\pi-\log(1-\pi)-(\x-\bm_1)\trans\Sigma^{-1}\bm_1-0.5\bm_1\trans\Sigma^{-1}\bm_1\}.
	\ese
	where $\bm_1=(-3,-3,-3,-3,0,\dots,0)\trans$.
	Then for the subjects with $R=1$, we generate the treatment assignment as
	\bse
	p(\x) = \pr(T=1\mid \x,R=1) = \expit(-0.5+0.125x_4),
	\ese
	in \underline{Case (i)}, and
	\bse
	p(\x) = \pr(T=1\mid \x,R=1) = 0.7,
	\ese
	in both \underline{Case (ii)} and \underline{Case (iii)}.
	Similar to Model 1, the potential outcomes are also generated by $Y_0=\mu^*_0(\X)+\epsilon_0$ and
	$Y_1=\epsilon_1$, where $\epsilon_0\sim N(0,1)$, $\epsilon_1\sim N(0,1/2)$ and $\mu_0^*(\cdot)$ is the same as in Model 1.
	In \underline{Model 2}, the outcome mean model $\mu_0(\x)$ will be correctly specified, and the model $p(\x)$ will also be correctly specified, whereas the model $\delta(\x)=\pi(\x)p(\x)$ will be misspecified in \underline{Case (i)} but correctly specified in \underline{Case (ii)} and \underline{Case (iii)}.
	
	In both Model 1 and Model 2, we consider a variety of combinations of $(N,d)$.
	In Model 1, the data generating mechanism results in about 40\% subjects from the primary study (the value $\pi$), and about 27\% subjects in the primary study who receive the treatment (the value $p$).
	In Model 2 Case (i) and Case (ii), the data generating mechanism results in about 40\% subjects in the primary study who receive the treatment (the value $p$), and we fix the sample size of the external controls as $m=1,000$.
	In Model 2 Case (iii), the data generating mechanism results in about 33\% subjects from the primary study (the value $\pi$).
	
	Based on 1,200 simulation replicates, we summarize the comparison results about the three estimators $\wh\theta\nv$, $\wh\theta\eff$, and $\wh\theta_{\wh{a}}$ in Table~\ref{tab:S1} for \underline{Model 1} (model $\mu_0(\x)$ misspecified, corresponds to (comp2) in Section~\ref{sec:naive}), in Table~\ref{tab:chi1} for \underline{Model 2 Case (i)} (model $\delta(\x)$ misspecified, corresponds to (comp1) in Section~\ref{sec:naive}), in Table~\ref{tab:chi3} for \underline{Model 2 Case (ii)} (no nuisance model misspecification), and in Table~\ref{tab:chi5} for \underline{Model 2 Case (iii)} (no nuisance model misspecification), respectively.
	For every estimator in each situation, we report the sample bias (Bias), the sample standard deviation (SD), the estimated standard error (SE), the 95\% coverage probability (CP), and the asymptotic relative efficiency (ARE) compared to the estimator $\wh\theta\nv$ which does not use external controls.
	Note that ARE is defined as the asymptotic variance of the estimator $\wh\theta\eff$ or $\wh\theta_{\wh{a}}$ divided by the asymptotic variance of $\wh\theta\nv$.
	Clearly ARE$<$1 indicates the corresponding estimator is more efficient than $\wh\theta\nv$.
	We follow Algorithm~\ref{alg} to compute the estimators.
	More specifically, the estimations of the propensity scores $p(\x)$, $\delta(\x)$, and the outcome regression $\mu_0(\x)$ are implemented using the \textbf{RCAL} and \textbf{glmnet} packages in \textsf{R}.
	In addition, all tuning parameters are selected via 5-fold cross-validation.
	
	In all simulations, Bias is close to zero indicating the asymptotic consistency of all the estimators, SD and SE are close to each other indicating the accuracy of the developed formulas of the asymptotic variance, and CP is also close to 95\% in many scenarios, though its performance might decline with the increasing dimensionality $d$ but improves with the increasing sample size.
	
	We are particularly interested in the efficiency comparisons among the three estimators.
	In Tables~\ref{tab:S1}, \ref{tab:chi3} and \ref{tab:chi5},
	the ARE is always smaller than 1 indicating that both $\wh\theta\eff$ and $\wh\theta_{\wh{a}}$ are more efficient than $\wh\theta\nv$.
	Further, the ARE of $\wh\theta_{\wh{a}}$ is even smaller than that of $\wh\theta\eff$ indicating that the estimator $\wh\theta_{\wh{a}}$ is even more efficient than $\wh\theta\eff$.
	Interestingly, in Table~\ref{tab:chi1} (model $\delta(\x)$ misspecified, corresponds to (comp1) in Section~\ref{sec:naive}), we encounter a phenomenon that the ARE of the estimator $\wh\theta\eff$ is greater than 1.
	This means $\wh\theta\eff$ is even less efficient than the estimator $\wh\theta\nv$---incorporating external controls might not obtain efficiency gain at all.
	Luckily, the ARE of the proposed estimator $\wh\theta_{\wh{a}}$ is strictly smaller than 1; it brings down the asymptotic variance and ensures the safe use of the external control data.
	
	In summary, $\wh\theta\eff$ could outperform $\wh\theta\nv$ theoretically when all the nuisance models are correctly specified.
	However, in real applications, estimation of the nuisance working models could be challenging, especially with high-dimensional covariates where correct model specification cannot be guaranteed.
	In such cases, we recommend using $\wh\theta_{\wh{a}}$ as it guarantees a safe and efficient estimate.

	\begin{table}[!htbp]
		\caption{
			Simulation results for the comparison of the three estimators $\wh\theta\nv$, $\wh\theta\eff$, and $\wh\theta_{\wh{a}}$ in \underline{Model 1}, with $n/N\approx 40\%$, $N\in\{1000,2000,3000\}$ and $d\in\{4,150,1000\}$.
			Note that ARE$<$1 indicates the corresponding estimator is more efficient than $\wh\theta\nv$.
		}
		\centering
		\begin{tabular}{rrrrrrrrrrr}
			\hline
			&       &                 \multicolumn{3}{c}{$d=4$}                 &                \multicolumn{3}{c}{$d=150$}                &               \multicolumn{3}{c}{$d=1000$}                \\
			$  n/N\approx 0.4$ &       & $\wh\theta\nv$ & $\wh\theta\eff$ & $\wh\theta_{\wh a}$ & $\wh\theta\nv$ & $\wh\theta\eff$ & $\wh\theta_{\wh a}$ & $\wh\theta\nv$ & $\wh\theta\eff$ & $\wh\theta_{\wh a}$ \\ \hline
			\multirow{5}{*}{$N=1000$} 		    &  Bias &           0.007 &            0.005 &                0.008 &           0.020 &            0.012 &                0.018 &           0.028 &            0.018 &                0.026 \\
			&    SD &           0.152 &            0.143 &                0.142 &           0.139 &            0.135 &                0.133 &           0.132 &            0.128 &                0.127 \\
			&    SE &           0.148 &            0.139 &                0.138 &           0.135 &            0.130 &                0.129 &           0.130 &            0.128 &                0.126 \\
			&    CP &           0.940 &            0.945 &                0.944 &           0.941 &            0.934 &                0.940 &           0.945 &            0.948 &                0.946 \\
			\cline{2-11}
			&  ARE  &            &            0.882 &                0.869 &            &            0.927 &                0.913 &            &            0.969 &                0.939 \\
			\hline
			\multirow{5}{*}{$N=2000$} 		    &  Bias &           0.002 &            0.002 &                0.003 &           0.015 &            0.010 &                0.014 &           0.020 &            0.014 &                0.018 \\
			&    SD &           0.111 &            0.105 &                0.105 &           0.103 &            0.098 &                0.098 &           0.094 &            0.092 &                0.091 \\
			&    SE &           0.105 &            0.099 &                0.099 &           0.098 &            0.094 &                0.094 &           0.096 &            0.093 &                0.092 \\
			&    CP &           0.930 &            0.936 &                0.933 &           0.932 &            0.929 &                0.932 &           0.948 &            0.950 &                0.949 \\
			\cline{2-11}
			&  ARE  &            &            0.889 &                0.889 &            &            0.920 &                0.920 &            &            0.938 &                0.918 \\
			\hline
			\multirow{5}{*}{$N=3000$} 		    &  Bias &           0.005 &            0.003 &                0.004 &           0.017 &            0.012 &                0.015 &           0.017 &            0.012 &                0.014 \\
			&    SD &           0.090 &            0.084 &                0.084 &           0.083 &            0.080 &                0.079 &           0.079 &            0.077 &                0.076 \\
			&    SE &           0.086 &            0.081 &                0.081 &           0.081 &            0.078 &                0.077 &           0.079 &            0.076 &                0.076 \\
			&    CP &           0.939 &            0.939 &                0.940 &           0.943 &            0.950 &                0.945 &           0.947 &            0.948 &                0.948 \\
			\cline{2-11}
			&  ARE  &            &            0.887 &                0.887 &            &            0.927 &                0.904 &            &            0.925 &                0.925 \\
			\hline
		\end{tabular}
		\label{tab:S1}
	\end{table}

	\begin{table}[!htbp]
		\caption{
			Simulation results for the comparison of the three estimators $\wh\theta\nv$, $\wh\theta\eff$, and $\wh\theta_{\wh{a}}$ in \underline{Model 2 Case (i)}, with $m=1000$, $n\in\{400,800,1200\}$ and $d\in\{4,100,1000\}$.
			Note that ARE$<$1 indicates the corresponding estimator is more efficient than $\wh\theta\nv$.
		}
		\centering
		\begin{tabular}{llrrrrrrrrr}
			\hline
			&       &                 \multicolumn{3}{c}{$d=4$}                 &                \multicolumn{3}{c}{$d=100$}                &               \multicolumn{3}{c}{$d=1000$}                \\
			$m=1,000$ &       & $\wh\theta\nv$ & $\wh\theta\eff$ & $\wh\theta_{\wh a}$ & $\wh\theta\nv$ & $\wh\theta\eff$ & $\wh\theta_{\wh a}$ & $\wh\theta\nv$ & $\wh\theta\eff$ & $\wh\theta_{\wh a}$ \\ \hline
			\multirow{5}{*}{$n=400$}           & Bias  &           0.001 &            0.010 &               -0.001 &           0.006 &            0.037 &                0.011 &           0.022 &            0.066 &                0.021 \\
			& SD    &           0.142 &            0.144 &                0.144 &           0.133 &            0.141 &                0.135 &           0.128 &            0.139 &                0.129 \\
			& SE    &           0.141 &            0.136 &                0.135 &           0.132 &            0.133 &                0.130 &           0.127 &            0.132 &                0.126 \\
			& CP    &           0.946 &            0.929 &                0.926 &           0.950 &            0.923 &                0.945 &           0.939 &            0.901 &                0.941 \\
			\cline{2-11}
			& ARE   &&0.930 &                0.917 &            &            1.015 &                0.970 &            &            1.080 &                0.984 \\
			
			\hline
			\multirow{5}{*}{$n=800$}          & Bias  &           0.001 &            0.001 &               -0.003 &           0.011 &            0.016 &                0.009 &           0.014 &            0.027 &                0.011 \\
			& SD    &           0.101 &            0.101 &                0.101 &           0.095 &            0.100 &                0.096 &           0.096 &            0.102 &                0.096 \\
			& SE    &           0.100 &            0.101 &                0.099 &           0.095 &            0.097 &                0.095 &           0.093 &            0.096 &                0.093 \\
			& CP    &           0.959 &            0.949 &                0.952 &           0.943 &            0.939 &                0.942 &           0.941 &            0.924 &                0.941 \\
			\cline{2-11}
			& ARE   &            &            1.020 &                0.980 &            &            1.042 &                1.000 &            &            1.065 &                1.000 \\
			\hline
			\multirow{5}{*}{$n=1200$} 	          & Bias  &           0.004 &            0.003 &                0.001 &           0.009 &            0.011 &                0.007 &           0.012 &            0.020 &                0.011 \\
			& SD    &           0.084 &            0.085 &                0.084 &           0.077 &            0.079 &                0.077 &           0.079 &            0.082 &                0.079 \\
			& SE    &           0.081 &            0.082 &                0.081 &           0.079 &            0.080 &                0.078 &           0.077 &            0.079 &                0.077 \\
			& CP    &           0.945 &            0.942 &                0.946 &           0.948 &            0.949 &                0.949 &           0.941 &            0.933 &                0.941 \\
			\cline{2-11}
			& ARE   &            &            1.025 &                1.000 &            &            1.025 &                0.975 &            &            1.053 &                1.000 \\
			\hline
		\end{tabular}
		\label{tab:chi1}
	\end{table}

	\begin{table}[!htbp]
		\caption{
			Simulation results for the comparison of the three estimators $\wh\theta\nv$, $\wh\theta\eff$, and $\wh\theta_{\wh{a}}$ in \underline{Model 2 Case (ii)}, with $m=1000$, $n\in\{400,800,1200\}$ and $d\in\{4,150,1000\}$.
			Note that ARE$<$1 indicates the corresponding estimator is more efficient than $\wh\theta\nv$.
		}
		\centering
		\begin{tabular}{llrrrrrrrrr}
			\hline
			&       &                 \multicolumn{3}{c}{$d=4$}                 &                \multicolumn{3}{c}{$d=150$}                &               \multicolumn{3}{c}{$d=1000$}                \\
			$m=1,000$ &       & $\wh\theta\nv$ & $\wh\theta\eff$ & $\wh\theta_{\wh a}$ & $\wh\theta\nv$ & $\wh\theta\eff$ & $\wh\theta_{\wh a}$ & $\wh\theta\nv$ & $\wh\theta\eff$ & $\wh\theta_{\wh a}$ \\ \hline
			\multirow{5}{*}{$n=400$} 	        & Bias  &          -0.004 &            0.000 &               -0.003 &           0.002 &            0.040 &                0.028 &          -0.001 &            0.073 &                0.051 \\
			& SD    &           0.128 &            0.132 &                0.126 &           0.127 &            0.130 &                0.126 &           0.130 &            0.131 &                0.128 \\
			& SE    &           0.129 &            0.121 &                0.118 &           0.119 &            0.110 &                0.109 &           0.113 &            0.104 &                0.102 \\
			& CP    &           0.951 &            0.920 &                0.933 &           0.926 &            0.888 &                0.894 &           0.899 &            0.835 &                0.863 \\
			\cline{2-11}
			& ARE   &            &            0.880 &                0.837 &            &            0.854 &                0.839 &            &            0.847 &                0.815 \\
			\hline
			\multirow{5}{*}{$n=800$} 		          & Bias  &          -0.002 &           -0.001 &               -0.003 &          -0.002 &            0.022 &                0.015 &          -0.004 &            0.032 &                0.020 \\
			& SD    &           0.091 &            0.093 &                0.089 &           0.090 &            0.093 &                0.090 &           0.089 &            0.091 &                0.088 \\
			& SE    &           0.092 &            0.089 &                0.087 &           0.087 &            0.082 &                0.081 &           0.084 &            0.080 &                0.079 \\
			& CP    &           0.949 &            0.941 &                0.942 &           0.940 &            0.906 &                0.921 &           0.931 &            0.892 &                0.913 \\
			\cline{2-11}
			& ARE  &            &            0.936 &                0.894 &            &            0.888 &                0.867 &            &            0.907 &                0.884 \\
			\hline
			\multirow{5}{*}{$n=1200$} 		          & Bias  &           0.001 &            0.003 &                0.001 &          -0.001 &            0.014 &                0.008 &          -0.001 &            0.023 &                0.015 \\
			& SD    &           0.076 &            0.075 &                0.073 &           0.074 &            0.075 &                0.073 &           0.070 &            0.073 &                0.071 \\
			& SE    &           0.075 &            0.073 &                0.071 &           0.072 &            0.069 &                0.068 &           0.070 &            0.067 &                0.067 \\
			& CP    &           0.947 &            0.947 &                0.949 &           0.936 &            0.924 &                0.926 &           0.952 &            0.915 &                0.926 \\
			\cline{2-11}
			& ARE   &            &            0.947 &                0.896 &            &            0.918 &                0.892 &            &            0.916 &                0.916 \\
			\hline
		\end{tabular}
		\label{tab:chi3}
	\end{table}
	
	\begin{table}[!htbp]
		\caption{
			Simulation results for the comparison of the three estimators $\wh\theta\nv$, $\wh\theta\eff$, and $\wh\theta_{\wh{a}}$ in \underline{Model 2 Case (iii)}, with $n/N\approx 33\%$, $n\in\{400,800,1200\}$ and $d\in\{4,150,1000\}$.
			Note that ARE$<$1 indicates the corresponding estimator is more efficient than $\wh\theta\nv$.}
		\centering
		\begin{tabular}{llrrrrrrrrr}
			\hline
			&       &                 \multicolumn{3}{c}{$d=4$}                 &                \multicolumn{3}{c}{$d=150$}                &               \multicolumn{3}{c}{$d=1000$}                \\
			$n/N=1/3$ &       & $\wh\theta\nv$ & $\wh\theta\eff$ & $\wh\theta_{\wh a}$ & $\wh\theta\nv$ & $\wh\theta\eff$ & $\wh\theta_{\wh a}$ & $\wh\theta\nv$ & $\wh\theta\eff$ & $\wh\theta_{\wh a}$ \\ \hline
			\multirow{5}{*}{$n=400$}                       & Bias  &          -0.003 &            0.002 &               -0.002 &          -0.004 &            0.036 &                0.023 &           0.007 &            0.081 &                0.059 \\
			& SD    &           0.131 &            0.134 &                0.129 &           0.129 &            0.132 &                0.127 &           0.126 &            0.122 &                0.123 \\
			& SE    &           0.129 &            0.121 &                0.119 &           0.118 &            0.111 &                0.109 &           0.113 &            0.104 &                0.102 \\
			& CP    &           0.939 &            0.922 &                0.920 &           0.922 &            0.886 &                0.907 &           0.919 &            0.816 &                0.852 \\
			\cline{2-11}
			& ARE   &            &            0.880 &                0.851 &            &            0.885 &                0.853 &            &            0.847 &                0.815 \\
			\hline
			\multirow{5}{*}{$n=800$}                 	        & Bias  &          -0.002 &           -0.001 &               -0.002 &          -0.002 &            0.021 &                0.014 &          -0.002 &            0.033 &                0.022 \\
			& SD    &           0.093 &            0.094 &                0.090 &           0.088 &            0.088 &                0.085 &           0.090 &            0.092 &                0.089 \\
			& SE    &           0.092 &            0.089 &                0.086 &           0.087 &            0.082 &                0.081 &           0.084 &            0.080 &                0.078 \\
			& CP    &           0.950 &            0.936 &                0.945 &           0.950 &            0.918 &                0.929 &           0.931 &            0.899 &                0.915 \\
			\cline{2-11}
			& ARE   &            &            0.936 &                0.874 &            &            0.888 &                0.867 &            &            0.907 &                0.862 \\
			\hline
			\multirow{5}{*}{$n=1200$}                 	        & Bias  &          -0.003 &           -0.002 &               -0.003 &          -0.005 &            0.009 &                0.003 &          -0.002 &            0.027 &                0.018 \\
			& SD    &           0.075 &            0.077 &                0.073 &           0.073 &            0.076 &                0.072 &           0.072 &            0.074 &                0.071 \\
			& SE    &           0.075 &            0.074 &                0.071 &           0.072 &            0.070 &                0.068 &           0.070 &            0.066 &                0.065 \\
			& CP    &           0.951 &            0.937 &                0.941 &           0.944 &            0.919 &                0.931 &           0.946 &            0.898 &                0.919 \\
			\cline{2-11}
			& ARE   &            &            0.973 &                0.896 &            &            0.945 &                0.892 &            &            0.889 &                0.862 \\
			\hline
		\end{tabular}
		\label{tab:chi5}
	\end{table}
	
		\section{Real Data Application}\label{sec:realdata}
		
		In this section, we demonstrate the proposed methods to a data set from the National Supported Work (NSW) study, which evaluates the impact of a job training program on future earning \citep{abadie2006large}.
		
		In our analysis, the NSW data set is a randomized controlled trial, and it consists of $289$ subjects including $111$ who received the job training program.
		Besides, our analysis also includes $554$ subjects from an external observational data set from the Current Population Survey (CPS), none of whom ever received the job training program.
		The job training program was implemented from 1975 to 1978.
		Our analysis focuses on individuals with no income in 1975.
		We adjust and analyze the income in 1978 after logarithmic transformation as the outcome, denoted as  $Y=\log (\mbox{Income in 1978}+1)$.
		The dataset contains six categorical covariates: age ($\leq 20$, $21-30$, $31-40$, $>40$), years of education ($\leq 6$, $7-9 $, $9-13 $, $>14$), Black ($1$ if black, $0$ otherwise), Hispanic ($1$ if Hispanic, $0$ otherwise), Marriage status ($1$ if married, $0$ otherwise), and whether you have a high school education ($1$ if you have no education, $0$ otherwise).
		These covariates are combined into $106$ distinct groups, treated as a $106$-dimensional covariate $\X$.
		
		We compare the performance of the three estimators $\wh\theta\nv$, $\wh\theta\eff$, and $\wh\theta_{\wh a}$ in Table~\ref{tab:real1}.
		The implementation follows Algorithm~\ref{alg}, with tuning parameters selected by 5-fold cross-validation.
		Table \ref{tab:real1} reveals that estimators incorporating external information ($\wh\theta\eff$ and $\wh \theta_{\wh a}$) exhibit small standard errors and significant $p$-values.
		In contrast, the estimator $\wh \theta\nv$ using only NSW data shows a large standard error and is not significant.
		This highlights the effectiveness of utilizing external information to enhance estimation efficiency.
		Moreover, the coefficient $\wh a=1.02$, very close to 1, indicating that
		the combined estimator $\wh \theta_{\wh a}$ is almost exactly the same as $\wh\theta\eff$; thus, using the estimator $\wh\theta\eff$ itself can already almost achieve the estimation efficiency.
		
		\begin{table}[tbp]
			\caption{Results from the real data application.}
			\centering
			\begin{tabular}{lrrr}
				\hline
				Estimator & Estimate & SE & p-value \\
				\hline
				$\wh \theta\nv$  & 0.780 & 0.49 & 0.1100 \\
				$\wh \theta\eff$  &  0.909 & 0.42 & 0.0313 \\
				$\wh \theta_{\wh a}$ &0.911 & 0.42 & 0.0308 \\
				\hline
			\end{tabular}\label{tab:real1}
		\end{table}

		\section{Discussion}\label{sec:disc}

		This work is motivated by the growing interest in incorporating external control data, typically easier to obtain from historical records or third-party sources, into randomized controlled trials (RCTs). 
		Our initial analysis reveals that standard doubly robust estimation can paradoxically lead to reduced efficiency compared to estimators that do not use external controls. 
		This counterintuitive outcome suggests that the naive inclusion of external controls may, in fact, be detrimental to estimation efficiency.
		
		To address this issue, we propose a novel doubly robust estimator that guarantees greater efficiency than the standard approach without external controls, even under model misspecification. 
		When all models are correctly specified, the proposed estimator coincides with the standard doubly robust estimator that incorporates external controls and achieves semiparametric efficiency. 
		The asymptotic theory developed here accommodates high-dimensional confounder settings, which are increasingly relevant given the growing prevalence of electronic health record data.
		
		For clarity of presentation, we adopt linear regression as the working model for the outcome mean and logistic regression as the working model for the propensity score in high-dimensional settings. 
		This choice allows us to present the key ideas transparently. 
		Nonetheless, the proposed framework can be generalized to more complex working models. 
		The corresponding technical developments lie beyond the scope of this paper and merit further investigation in future research.

		\section*{Supplement}
		
		The supplement includes all the technical details.
		
		\section*{Acknowledgment}
		
		The research is supported in part by NSF (DMS 1953526, 2122074, 2310942), NIH (R01DC021431) and the American Family Funding Initiative of UW-Madison.

		\section*{Conflict of Interest}

		The authors report there are no competing interests to declare.
		
		\bibliographystyle{agsm}
		\bibliography{refSSLcontrol.bib}

@book{buhlmann2011statistics,
  title={Statistics for High-Dimensional Data: Methods, Theory and Applications},
  author={B{\"u}hlmann, Peter and van de Geer, Sara},
  year={2011},
  publisher={Springer Science \& Business Media}
}

@article{farrell2015robust,
	title={Robust inference on average treatment effects with possibly more covariates than observations},
	author={Farrell, Max H},
	journal={Journal of Econometrics},
	volume={189},
	number={1},
	pages={1--23},
	year={2015},
	publisher={Elsevier}
}

@article{li2021improving,
  title={Improving efficiency of inference in clinical trials with external control data},
  author={Li, Xinyu and Miao, Wang and Lu, Fang and Zhou, Xiao-Hua},
  journal={Biometrics},
  volume={79},
  number={1},
  pages={394--403},
  year={2023},
  publisher={Wiley Online Library}
}

@article{schwartz2023harmonized,
  title={Harmonized estimation of subgroup-specific treatment effects in randomized trials: The use of external control data},
  author={Schwartz, Daniel and Saha, Riddhiman and Ventz, Steffen and Trippa, Lorenzo},
  journal={arXiv preprint arXiv:2308.05073},
  year={2023}
}

@article{liao2023prognostic,
  title={Prognostic adjustment with efficient estimators to unbiasedly leverage historical data in randomized trials},
  author={Liao, Lauren D and H{\o}jbjerre-Frandsen, Emilie and Hubbard, Alan E and Schuler, Alejandro},
  journal={arXiv preprint arXiv:2305.19180},
  year={2023}
}

@article{valancius2024causal,
  title={A causal inference framework for leveraging external controls in hybrid trials},
  author={Valancius, Michael and Pang, Herbert and Zhu, Jiawen and Cole, Stephen R and Jonsson Funk, Michele and Kosorok, Michael R},
  journal={Biometrics},
  volume={80},
  number={4},
  pages={ujae095},
  year={2024},
  publisher={Oxford University Press}
}

@article{zhu2024enhancing,
  title={Enhancing Statistical Validity and Power in Hybrid Controlled Trials: A Randomization Inference Approach with Conformal Selective Borrowing},
  author={Zhu, Ke and Yang, Shu and Wang, Xiaofei},
  journal={arXiv preprint arXiv:2410.11713},
  year={2024}
}

@article{gao2023integrating,
  title={Integrating Randomized Placebo-Controlled Trial Data with External Controls: A Semiparametric Approach with Selective Borrowing},
  author={Gao, Chenyin and Yang, Shu and Shan, Mingyang and Ye, Wenyu and Lipkovich, Ilya and Faries, Douglas},
  journal={arXiv preprint arXiv:2306.16642},
  year={2023}
}

@article{yang2023sam,
  title={SAM: Self-adapting mixture prior to dynamically borrow information from historical data in clinical trials},
  author={Yang, Peng and Zhao, Yuansong and Nie, Lei and Vallejo, Jonathon and Yuan, Ying},
  journal={Biometrics},
  volume={79},
  number={4},
  pages={2857--2868},
  year={2023},
  publisher={Oxford University Press}
}

@article{ohigashi2024nonparametric,
  title={Nonparametric Bayesian approach for dynamic borrowing of historical control data},
  author={Ohigashi, Tomohiro and Maruo, Kazushi and Sozu, Takashi and Gosho, Masahiko},
  journal={arXiv preprint arXiv:2411.11675},
  year={2024}
}

@article{neyman1923applications,
  title={Sur les applications de la thar des probabilities aux experiences Agaricales: Essay des principle. Excerpts reprinted (1990) in English},
  author={Neyman, Jerzy},
  journal={Statistical Science},
  volume={5},
  number={463-472},
  pages={4},
  year={1923}
}

@article{rubin1974estimating,
  title={Estimating causal effects of treatments in randomized and nonrandomized studies.},
  author={Rubin, Donald B},
  journal={Journal of Educational Psychology},
  volume={66},
  number={5},
  pages={688},
  year={1974},
  publisher={American Psychological Association}
}

@book{bickel1993efficient,
  title={Efficient and Adaptive Estimation for Semiparametric Models},
  author={Bickel, Peter J and Klaassen, J and Ritov, YA'Acov and Wellner, Jon A},
  year={1993},
  publisher={Johns Hopkins University Press Baltimore}
}

@book{tsiatis2006semiparametric,
  title={Semiparametric Theory and Missing Data},
  author={Tsiatis, Anastasios A},
  year={2006},
  publisher={New York: Springer}
}

@article{heckmansmith1995,
	title={Assessing the case for social experiments},
	author={Heckman, James J and Smith, Jeffrey A},
	journal={Journal of Economic Perspectives},
	volume={9},
	number={2},
	pages={85--110},
	year={1995}
}

@article{wang2017g,
	title={G-computation of average treatment effects on the treated and the untreated},
	author={Wang, Aolin and Nianogo, Roch A and Arah, Onyebuchi A},
	journal={BMC Medical Research Methodology},
	volume={17},
	number={1},
	pages={1--5},
	year={2017},
	publisher={Springer}
}

@article{heckman2001policy,
	title={Policy-relevant treatment effects},
	author={Heckman, James J and Vytlacil, Edward},
	journal={American Economic Review},
	volume={91},
	number={2},
	pages={107--111},
	year={2001}
}

@article{abadie2006large,
	title={Large sample properties of matching estimators for average treatment effects},
	author={Abadie, Alberto and Imbens, Guido W},
	journal={Econometrica},
	volume={74},
	number={1},
	pages={235--267},
	year={2006},
	publisher={Wiley Online Library}
}

@book{imbens2015causal,
  title={Causal Inference in Statistics, Social, and Biomedical Sciences},
  author={Imbens, Guido W and Rubin, Donald B},
  year={2015},
  publisher={Cambridge University Press}
}

@book{hernan_robins_2022, place={Boca Raton}, title={Causal Inference: What If}, isbn={9781420076165}, publisher={Taylor \& Francis}, author={Hernan, Miguel A and Robins, James M}, year={2022}}

@book{Morgan_Winship_2014, place={Cambridge}, title={Counterfactuals and Causal Inference}, DOI={https://doi.org/10.1017/CBO9781107587991}, publisher={Cambridge University Press},
	edition={2nd}, author={Morgan, Stephen L and Winship, Christopher}, year={2014}}

@article{ghosh2022doubly,
  title={Doubly robust semiparametric inference using regularized calibrated estimation with high-dimensional data},
  author={Ghosh, Satyajit and Tan, Zhiqiang},
  journal={Bernoulli},
  volume={28},
  number={3},
  pages={1675--1703},
  year={2022},
  publisher={Bernoulli Society for Mathematical Statistics and Probability}
}

@article{tan2020model,
  title={Model-assisted inference for treatment effects using regularized calibrated estimation with high-dimensional data},
  author={Tan, Zhiqiang},
  journal={Annals of Statistics},
  volume={48},
  number={2},
  pages={811--837},
  year={2020},
  publisher={Institute of Mathematical Statistics}
}

@article{ning2020robust,
  title={Robust estimation of causal effects via a high-dimensional covariate balancing propensity score},
  author={Ning, Yang and Sida, Peng and Imai, Kosuke},
  journal={Biometrika},
  volume={107},
  number={3},
  pages={533--554},
  year={2020},
  publisher={Oxford University Press}
}

@article{tan2020regularized,
	title={Regularized calibrated estimation of propensity scores with model misspecification and high-dimensional data},
	author={Tan, Zhiqiang},
	journal={Biometrika},
	volume={107},
	number={1},
	pages={137--158},
	year={2020},
	publisher={Oxford University Press}
}

@article{belloni2018uniformly,
  title={Uniformly valid post-regularization confidence regions for many functional parameters in z-estimation framework},
  author={Belloni, Alexandre and Chernozhukov, Victor and Chetverikov, Denis and Wei, Ying},
  journal={The Annals of Statistics},
  volume={46},
  number={6B},
  pages={3643--3675},
  year={2018},
  publisher={Institute of Mathematical Statistics}
}

@article{boukhris2016antidepressant,
  title={Antidepressant use during pregnancy and the risk of autism spectrum disorder in children},
  author={Boukhris, Takoua and Sheehy, Odile and Mottron, Laurent and B{\'e}rard, Anick},
  journal={JAMA pediatrics},
  volume={170},
  number={2},
  pages={117--124},
  year={2016},
  publisher={American Medical Association}
}

@article{zhang2021dynamic,
  title={Dynamic treatment effects: high-dimensional inference under model misspecification},
  author={Zhang, Yuqian and Ji, Weijie and Bradic, Jelena},
  journal={arXiv preprint arXiv:2111.06818},
  year={2021}
}

@article{athey2018approximate,
  title={Approximate residual balancing: debiased inference of average treatment effects in high dimensions},
  author={Athey, Susan and Imbens, Guido W and Wager, Stefan},
  journal={Journal of the Royal Statistical Society Series B: Statistical Methodology},
  volume={80},
  number={4},
  pages={597--623},
  year={2018},
  publisher={Oxford University Press}
}

@article{sun2022high,
  title={High-dimensional model-assisted inference for local average treatment effects with instrumental variables},
  author={Sun, Baoluo and Tan, Zhiqiang},
  journal={Journal of Business \& Economic Statistics},
  volume={40},
  number={4},
  pages={1732--1744},
  year={2022},
  publisher={Taylor \& Francis}
}

@article{baybutt2023doubly,
  title={Doubly-Robust Inference for Conditional Average Treatment Effects with High-Dimensional Controls},
  author={Baybutt, Adam and Navjeevan, Manu},
  journal={arXiv preprint arXiv:2301.06283},
  year={2023}
}

@article{smucler2019unifying,
  title={A unifying approach for doubly-robust $\ell_1$ regularized estimation of causal contrasts},
  author={Smucler, Ezequiel and Rotnitzky, Andrea and Robins, James M},
  journal={arXiv preprint arXiv:1904.03737},
  year={2019}
}

@article{wu2021model,
  title={Model-assisted inference for covariate-specific treatment effects with high-dimensional data},
  author={Wu, Peng and Tan, Zhiqiang and Hu, Wenjie and Zhou, Xiao-Hua},
  journal={arXiv preprint arXiv:2105.11362},
  year={2021}
}

\end{document}